\documentclass{elsart}
\usepackage[]{epsfig}
\begin{document}

\begin{frontmatter}

\title{Regular black holes in UV self-complete quantum gravity}

\author{Euro Spallucci\thanksref{infn}}
\thanks[infn]{e-mail address: spallucci@ts.infn.it }
\address{Dipartimento di Fisica Teorica, Universit\`a di Trieste
and INFN, Sezione di Trieste, Italy}

\author{Stefano Ansoldi\thanksref{s}}
\thanks[s]{e-mail address: stefano.ansoldi@dimi.uniud.it}
\address{International Center for Relativistic Astrophysics (ICRA)
and INFN, Sezione di Trieste,
and Dipartimento di Matematica e Informatica, Universit\`{a} di Udine, Italy}

\begin{abstract}
In this letter we investigate the role of regular (curvature singularity-free)
black holes in the framework of UV self-complete quantum gravity.
The existence of a minimal length, shielding the trans-Planckian regime
to any physical probe, is self-consistently included into the black hole
probe itself. In this way we obtain to slightly shift the barrier below the 
Planck Length, with the UV self-complete scenario self-consistently confirmed.
\end{abstract}
\end{frontmatter}

\section{Introduction}

The nature of  space and time at the Planck scale is a longstanding
argument of debate. Fluctuations in both geometry and topology are expected to
become so violent  to disrupt the very fabric
of the spacetime manifold. The term ``spacetime foam'' is frequently
used to portray this kind of gravitational quantum vacuum \cite{Wheeler:1998vs}.
Any candidate theory of quantum gravity has to address this problem
and provide some information about trans-Planckian physics, whatever
it is. Even if String Theory is not yet a fully accomplished Unified
Theory of Everything, it provides to day the most powerful framework
to address quantum gravity problems.
The price to pay for that is to dismiss the idea of ``point-like''
building blocks of matter in favor of one-dimensional, Planck size,
fundamental objects. Unfortunately, the extended nature of (super) strings 
makes them unable to probe the trans-Planckian regime: as opposed to 
hypothetical point-like
objects, increasing the energy is not enough to make them shorter and shorter; 
as more
and more excitation modes are switched-on, the string elongates
\cite{Amati:1988tn} bouncing back to a long-distance regime
\footnote{For a different approach to a string induced minimal length 
see \cite{Fontanini:2005ik,Spallucci:2005bm}.}.
A quite different approach to the problem has been recently
proposed by Dvali and collaborators in a series of papers
\cite{Dvali:2010bf,Dvali:2010ue}, where String Theory is not
explicitly involved. We shall comment this feature in the conclusions.

\begin{figure}[ht!]
\begin{center}
\includegraphics[width=10.5cm,angle=0]{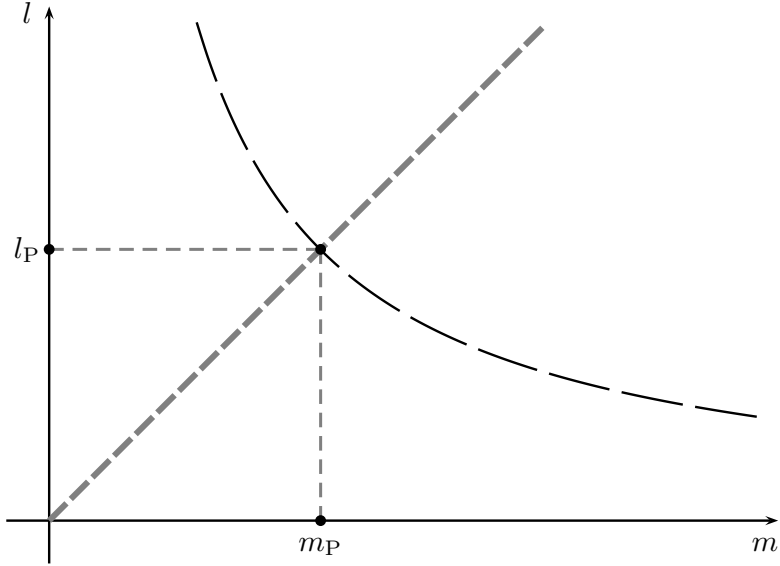}
\caption{\label{Planck} The hyperbola represents the Compton wavelength
of a ``particle'' of mass $m$. The straight line shows the linear
increase of the  Schwarzschild radius with respect to the mass of a black
hole.  The intersection between the two curves defines the
Planck Length, $l _{\mathrm{P}}$, and the Planck Mass, $m _{\mathrm{P}}$. }
\end{center}
\end{figure}

The general wisdom says that there is no self-consistent
way to quantize gravity in the framework of ``point-like''
quantum field theory because in the foamy Planckian phase quantum
fluctuations are out of control, and predictive power is lost even in 
super-gravity
models. Against this background, Dvali proposed a clever way
to by-pass such a problem, by pointing-out the existence of
a ``black hole barrier''  shielding the trans-Planckian regime
to any physical probe. In a nutshell,
gravity regularizes itself because of its unique ability to collapse
high enough energy concentrations into black holes, with linear dimension
increasing with energy, and not vice-versa.  Thus, any point-like
probe turns into a black hole when boosted to a ``critical
energy'' $ -s_\ast=\hbar c/2G_N $. Any further mass-energy increase
reverses the Lorentz contraction in a sort of
\emph{Schwarzschild dilation} of the gravitational radius
$R_s= 2G_N\sqrt{-s}/c^2$. The effect of gravity is to \emph{shield}
the deep-UV region behind the curtain of an event horizon
(See fig.~\ref{Planck}).

The far reaching conclusion of this simple reasoning is that,
contrary to any current wisdom,  the quantum gravity
trans-Planckian regime could be dominated by ``classical'', infra-red, field
configurations. This result is reminiscent of $T$-Duality in String Theory,
where a stringy probe cannot distinguish a length scale $L$ from a length
$\alpha^\prime/L$. Thus, $\sqrt{\alpha^\prime}$ is the ultimate accessible
distance to a stringy object. Keeping this in mind, a unique, and often 
overlooked, black hole property is to provide an ideal bridge between micro 
and macro physics \cite{Chapline:2000en,Laughlin:2003yh,Kosyakov:2007zv}. 
Indeed, whatever the radius of the horizon, a black hole
is always  a ``point-like'' object, in the sense that the whole mass
is packed inside an arbitrarily small region around the origin (classically
the mass is collapsed into a single point). A black hole can be seen as
a self-gravitating particle, and the infinite self-force the field
applies to its own point-like source translates into the presence of a
``curvature singularity''. From this point of view, a $1\,\mathrm{gr}$ 
mass black hole
can be seen either as an infra-red field configuration of radius 
$10^5\, l_{\mathrm{P}}$, or
a trans-Planckian test-particle with energy   $10^5\, E_{\mathrm{P}}$.

Then, in the scenario briefly discussed above, the \emph{minimal}, physically
meaningful, length turns out to be the \emph{Planck length} $l_{\mathrm{P}}$, 
which
is defined as the \emph{cross-over} point between the
Schwarzschild radius of a mass $m$ black hole and the Compton wavelength
of a particle with the same mass:
\begin{equation}
l_{\mathrm{P}}\equiv (R_s)_{min}  = 
\sqrt{\frac{2\hbar G_N}{c^3}} \label{lplanck}
\end{equation}
Any distance $d<l_{\mathrm{P}} $ has no physical meaning being shielded
by the horizon. Fig.~\ref{Planck}
is a portrait of an elementary objects ``\emph{phase space}''.
Light objects with $m < m_{\mathrm{P}}$ are what we colloquially call 
``particles''.
Their linear dimension is defined by the Compton wavelength encoding the
quantum mechanical nature of a microscopic object. On the other hand, heavy 
objects
with $m > m_{\mathrm{P}}$ are gravity-dominated and they look like
\emph{classical} black holes of linear size $R_s$.
The Planck scale represents the \emph{critical point} where Quantum Mechanics 
intersects
General Relativity and the Compton wavelength is ``swallowed''
by a ``classical'' black holes (the term ``classical'' means
``solution of the Einstein equations'', and is not referring to the actual
size of the object\footnote{To appreciate the specific meaning of ``classical''
in this framework, it may be useful to recall an analogy with Yang-Mills 
``instantons''.
Also in this case one talks of ``classical solutions'', even if such field
configurations are confined to microscopic scale. With this analogy in mind,
one can say that instantons play an essential role in non-perturbative
Yang-Mills theory, and black holes control gravity in the trans-Planckian
regime. In both cases, the dynamics of the theory is described in terms
of classical field configurations instead of particle-like
excitations  \cite{Dvali:2010jz,Dvali:2010ns}.}).
The existence of a black hole barrier follows as a necessary consequence
from the purely attractive character of gravity and is instrumental to the
realization of the UV self-complete scenario. A possible critical remark to
this scenario that has been raised in the literature, is that a Planckian
black hole is highly unstable with respect to Hawking evaporation. Thus,
it is conceivable that a Planckian probe will disintegrate, soon after
its formation, into a burst of thermal radiation. While emitting Hawking 
radiation
the black hole will shrink to smaller and smaller size. Then,
in principle, a decaying black hole \emph{can} probe
distances smaller than the Planck length, at least during the final phase
of its evaporation process. More precisely, one should say
that the structure of the probe in these extreme conditions is unknown:
maybe a transition to some excited string state could occur
\cite{Horowitz:1996nw}, and the whole self-completeness argument would require
to be adapted to this different situation.

The purpose of this letter is to provide an answer to this criticism.
The root of the problem can be traced back to the fact that
in the Schwarzschild geometry
there is no lower bound to the radius of the black hole horizon
during the evaporation process.  This is, again, a consequence
of the possibility to consider the source of the field concentrated into an
arbitrary small volume. On the other hand, if a minimal distance exists
point-like sources have no physical meaning.  From standard
quantum mechanics we know that "point-like", classical, particles can at most 
be represented by optimal  localization, or minimal uncertainty, position states.
In a recent series of papers
\cite{Nicolini06,Nicolini:2005de,Ansoldi07,Spallucci:2008ez,Nicolini:2009gw,Smailagic:2010nv}
we introduced  this idea in General Relativity and found
black hole solutions generated by a minimal width Gaussian distribution of matter.
For the reader's convenience, we list below the main features of these objects.\\
i) They are curvature singularity free. This is a straightforward consequence of spreading
the source over a finite volume. The arbitrary large curvature region close to the origin
is turned into a de~Sitter vacuum core with finite curvature.\\
ii) They admit an extremal, degenerate, configuration even in the neutral, non-rotating case.
The presence of both an inner (Cauchy) horizon and an outer (Killing) horizon is a 
characteristic
feature of this regular solutions\footnote{The stability of the Cauchy
horizon is an open issue which is currently under investigation 
\cite{Batic:2010vm,Brown:2010cr}.
In any case, this discussion is not relevant to the problem we are discussing in this work.}.
\\
iii) The Hawking temperature is bounded from above and vanishes
for the extremal configuration. The heat capacity is positive in the small black
hole phase, making these solutions thermodynamically stable.
\\
iv) A detailed investigation of the quantum properties of these objects, in relation to 
production and decay at LHC can be found in \cite{Rizzo:2006zb,Casadio:2008qy,Gingrich:2010ed}.
\\
In what follows we will see how it is possible to make \emph{self-consistent}
the UV self-completeness proposal by taking into account the
existence of a minimal length in the black hole probe itself.
The advantage of this approach is that the minimal size black hole
is a zero Hawking temperature, stable, extremal configuration,
which will evade the above mentioned criticism.

\section{Regular Schwarzschild black hole}

Black hole type solutions of the Einstein field equations are plagued by
the presence of curvature singularities, where tidal forces arbitrarily blow up.
From a physical point of view, no measurable quantity can become infinite.
Indeed, the presence of a singularity cannot be seen as a
``physical'' effect, rather it sounds like a warning that we are pushing a
classical theory, i.e. General Relativity, where it stops to be effective and
loses its predictive power. A possible cure to the ``singularity
sickness'' is suggested by non-commutative geometry, where manifold
fluctuations make it impossible to measure
lengths shorter than a minimal length $\sqrt{\theta}$. The parameter $\theta$
is a measure of how much non-commuting coordinates deviate from their
classical, commuting, counterparts. In a series of papers we introduced
a phenomenological approach where the key feature of non-commuting geometry,
i.e. the existence of a minimal length, is encoded into Einstein equations
by re-modelling matter sources in terms of minimal
width Gaussian distributions. For more details we refer the reader to the
original papers 
\cite{Ansoldi:2008jw,Nicolini:2008aj,Banerjee:2009gr,Modesto:2010uh}. 
We would only like to comment about the sensitivity of the solution
with respect to the choice of the source. Regular black holes can be
obtained both by coupling gravity to non-linear electrodynamics 
\cite{AyonBeato:1998ub,AyonBeato:1999ec,AyonBeato:1999rg}, 
and by enginnering appropriate sources, e.g. \cite{Hayward:2005gi,Shankaranarayanan:2003qm}
(~for a general review about this topic, see \cite{Ansoldi:2008jw}~).
In our case, the gaussian form of the matter distribution is not a choice
but an exact result recovered from the underlying non-commutative geometry.
Strictly speaking, we could extend the gaussian distribution to a Maxwell-like
form, i.e.  $\rho_G\left(\, r\,\right) \longrightarrow r^n\, \rho_G\left(\, r\,\right) $,
 $n\ge 0$ integer,
without spoiling the regularity of the black hole solution.  The physical
difference is clear, we replace a massive droplet source with an hollow shell of matter.
From the geometrical side, the inner deSitter core will be replaced by
flat Minkowski central region. All the other appealing features of the black hole
solution are preserved.\\
The simplest solution
of the modified Einstein equations is the so-called ``non-commutative''
Schwarzschild metric
\begin{eqnarray}
ds^2  & = & -f(r) dt^2 + f^{-1}(r) dr^2 + r^2 d\Omega_2 ^{2}\ ,
\nonumber  \\
d \Omega_2 ^{2} & \equiv & d\theta^2 + \sin^2\theta\, d\phi^2
\ , \quad
(c=1\ , G_N=1)
\label{nss} \\
f(r) & = & 1 -\frac{4M}{\sqrt\pi\, r}\gamma\left(\, \frac{3}{2}\ ,
\frac{r^2}{4\theta}\,\right)\ ,\quad
\gamma\left(\, \frac{3}{2}\ ,\frac{r^2}{4\theta}\,\right) =
\int_0^{r^2/4\theta} dt\, t^{1/2}\, e^{-t} \ ,
\nonumber
\end{eqnarray}
where, $\sqrt{\theta}$ is the width of the Gaussian mass-energy distribution
of the source.
Expanding $f\left(r\right)$ near the origin, one sees that
the central curvature singularity is replaced by a de~Sitter vacuum core
characterized by an effective cosmological constant
$\Lambda_{\mathrm{eff.}}=M/\left(\sqrt\pi\,\theta^{3/2}\right)$. The line element
(\ref{nss}) smoothly interpolates between the de~Sitter geometry at short
distance, i.e. $r \ll \sqrt\theta $, and the Schwarzschild metric at large distance
$r \gg \sqrt\theta$. Some cautionary remark about the short distance limit is due.
This is the range  where our effective description breaks down and the very
concept of smooth space-time loses its meaning. However, through the
looking glass of gravity a  non-commutative fluctuating manifold is filtered
into a non-trivial ``vacuum'' of de~Sitter type.

\begin{figure}[ht!]
\begin{center}
\includegraphics[width=10.5cm,angle=0]{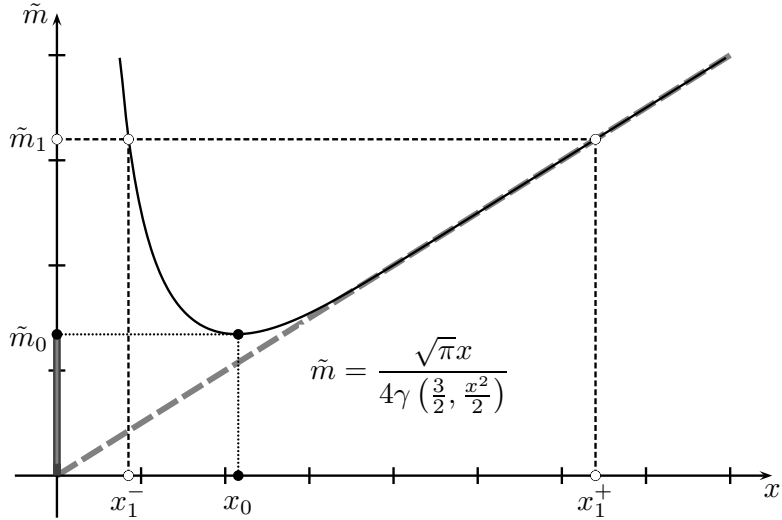}
\caption{\label{hreg} This is the plot of Eq.(\ref{horizons})
in terms of rescaled variables $x\equiv r_H/\sqrt{2\theta}$,
$\tilde{m}\equiv MG_N/\sqrt{2\theta}$. $\tilde{m}_0$, $x_0$ are the mass
and radius of the ``extremal'' black hole configuration.
If $\tilde{m}>\tilde{m}_0$ we have a non-degenerate black hole
with event horizon of radius $x_+$ and inner Cauchy horizon of radius $x_-$.
For $\tilde{m}<\tilde{m}_0 $ we have a particle-like object with no
horizons. }
\end{center}
\end{figure}

In the intermediate distance range non-standard black hole configurations can be realized
above a certain mass threshold. Let us consider the zeros of the metric function,  
$f(r_H)=0$ and plot the
total mass-energy $M$ as a function of the Schwarzschild radius $r_H$ (see Fig.~\ref{hreg})
\begin{equation}
M=\frac{\sqrt\pi}{4}
\frac{r_H}{\gamma\left(\, 3/2\ , r^2_H/4\theta\,\right)}
\label{horizons}
\end{equation}

Even a neutral, non-spinning, object of mass $M_1> M_0$ is a black hole with an outer 
(Schwarzschild)
horizon of radius $r_1^+$ and an inner (Cauchy) horizon of radius $r_1^-$.
As the mass decreases towards $M_0$ the two horizons merge into a single,
degenerate, null surface, with $r_H=r_0$. This is an \emph{extremal} black hole.
For lower masses there are no more horizons and the object is a regular,
particle-like, lump of matter.
The presence of two horizons and the existence of an extremal configuration
make the thermodynamic behavior of this uncharged, non-spinning, black hole quite
similar to the thermal evolution of a standard, charged, Reissner-Nordstrom
black hole.  The Hawking temperature is bounded from above and vanishing
as the extremal configuration is reached. This is the so-called ``scram-phase''
\cite{Nicolini:2008aj} leading to a stable massive remnant in the form of
a degenerate extremal black hole. This end-point configuration  is the
most relevant one in  the framework of self-complete quantum gravity as it
provides us the smallest probe we can think of. Let us give a closer look
to the extremal black hole represented by the minimum of the curve in
Fig.~\ref{hreg}. The minimum is characterized by
\begin{equation}
f(r_0) = 0
\label{eq:sis001}
\end{equation}
\begin{equation}
\left(\, \frac{dM}{dr_H}\,\right)_{r_H=r_0} = 0
\quad \Rightarrow \quad
r_0^3=4\theta^{3/2}\gamma\left(\, \frac{3}{2}\ ,\frac{r^2_0}{4\theta}\,\right)
e^{r_0^2/4\theta}
.
\label{eq:sis002}
\end{equation}
Notice that in terms of the length unit $\sqrt{\theta}$ the horizon
curve becomes $\theta$-independent while the Compton hyperbola can
be shifted by varying the value of $\theta$:
\[
\frac{l _{\mathrm{C}}}{\sqrt{2 \theta}} = \frac{\hbar}{2 \theta ( m / \sqrt{2 \theta} )}
= \frac{l _{\mathrm{P}} ^{2}}{4 \theta} \cdot \frac{1}{m / \sqrt{2 \theta}}
 = \frac{l _{\mathrm{P}} ^{2}}{4 \theta} \cdot \frac{1}{\tilde{m}} .
\]
Thus, having the rescaled horizon
curve fixed and the rescaled Compton wavelength freely adjustable, it is consistent
to look for the value of $\theta$ allowing an intersection
point between the two curves for the values of mass and event horizon radius that
define an extremal configuration. This peculiar crossing point is obtained
for $\sqrt{\theta} \simeq l_{\mathrm{P}}/3.393$,
\begin{figure}[ht!]
\begin{center}
\includegraphics[width=10.5cm,angle=0]{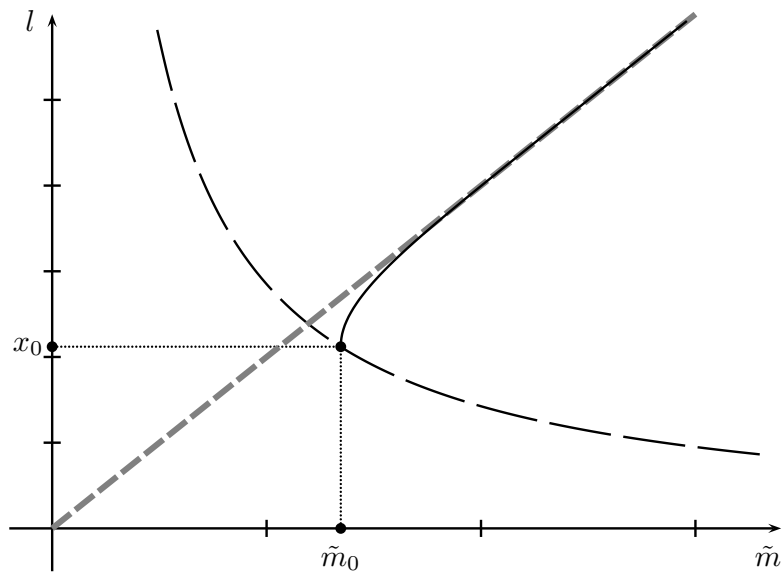}
\caption{\label{regular} The dashed curve is the rescaled Compton hyperbola
for the critical value $ \sqrt{\theta} \simeq l_{\mathrm{P}}/3.393 $. The continuous curve
is the outer horizon branch of (\ref{horizons}).}
\end{center}
\end{figure}
A relationship between the minimal length $r _{0}$ and the Planck length not involving
$M _{0}$ can then be obtained combining (\ref{eq:sis001}), (\ref{eq:sis002}) and
$r _{0} = l _{\mathrm{P}} ^{2} / (2 M _{0})$
\begin{equation}
    r _{0} ^{3}
    =
    \frac{\sqrt{\pi} r _{0}}{M _{0}} \theta ^{3/2} e ^{r_{0} ^{2} / (4 \theta)}
    =
    2 \sqrt{\pi} \left( \frac{\sqrt{\theta}}{l _{\mathrm{P}}} \right)^{3} \, r _{0} ^{2} \,
    e ^{r_{0} ^{2} / (4 \theta)} l _{\mathrm{P}}
\label{new}
\end{equation}
so that
\[
L_\ast \stackrel{\mathrm{def.}}{=} r _{0} \simeq 2\sqrt\pi \frac{9.8138}{(3.393)^3}\,
l_{\mathrm{P}}
\simeq 0.891\,l _{\mathrm{P}}
\quad \mathrm{and} \quad
M_\ast \stackrel{\mathrm{def.}}{=} M _{0} = \frac{\hbar}{L _{\ast}}
\]
are the new values for the ``Planck'' length and mass.
Thus, the black hole barrier is just slightly shifted below
the Planck scale and the UV self-completeness
scenario is self-consistently preserved.

\section{Conclusions}

We  conclude this note by pointing out some interesting connections among
self-complete quantum gravity, string theory, non-commutative
geometry, regular back holes and un-particles.

The very concept of point-particle is only a low energy approximation
for a one-dimensional string, and the naive idea that shorter and shorter
length scales can be probed by injecting more ad more energy into the probe
breaks down at the string scale  $l_s=\sqrt{\alpha^\prime}$.
To make contact with the UV self-complete scenario we recall the
Correspondence Principle for Black holes and Strings \cite{Horowitz:1996nw}. In
\cite{Susskind:1993ws} Susskind
suggested that there exists a one-to-one correspondence between
Schwarzschild black holes and fundamental string states. The argument
follows from the fact that in the strong coupling regime the size
of an highly excited string is less than its Schwarzschild radius.
On the other hand, the interest for non-commutative geometry was
boosted in the high energy physics community by the recognition that
spacetime coordinates
turn into non-commuting objects as an effect of string-$D$-brane coupling
in the presence of a Nouveau-Schwartz background field \cite{Witten:1985cc,Seiberg:1999vs}.
Uncertainty in the localization of any physical event, near and beyond a certain length 
scale
$l_{NC}=\sqrt{\theta}$, becomes an unavoidable feature of any physical theory.
We encoded this intrinsic limit into our regular black hole solution
by smearing the central curvature singularity, or mass-energy density, into
a minimal width Gaussian distribution.

Finally, self-complete quantum gravity
provides a different view  of the minimal distance which
can be probed in a \emph{gendanken} high energy experiment as the radius
of a thermodynamically stable, extremal, regular black hole. Our
self-consistent approach allows to push the black hole barrier slightly below
the Planck length, but it is still there. Is this the end of the story?

A couple of years ago Georgi introduced a possible new sector of the elementary particle
Standard model, where scale invariance is realized in the form of a continuous mass spectrum
\cite{Georgi:2007ek,Georgi:2007si}.  The new objects have been called \emph{un-particles}
to distinguish them from ordinary matter. The interactions between particles and un-particles
introduce an entire new phenomenology to be, hopefully,  tested at LHC.
As far as gravity is concerned, un-gravitons\footnote{The effective actions for various
unparticle fields have been discussed in \cite{Gaete:2008wg,Gaete:2008aj}.} lead to  
deviations from the Newton law \cite{Goldberg:2007tt}
which turn, at the non-perturbative level, into modifications of the Schwarzschild
geometry \cite{Mureika:2007nc,Mureika:2008dx,Mureika:2010je,Gaete:2010sp}.
The un-graviton modified metric results to be formally equivalent to the line element in the
presence of \emph{fractal}  extra dimensions.  The non-trivial way scale invariance is 
realized
in the un-particle sector seems to be the key to access a new \emph{fractal phase} of 
space-time
geometry \cite{Calcagni:2010bj,Calcagni:2009kc,Modesto:2009qc,Nicolini:2010dj,Calcagni:2010pa}.
A recent analysis  of high energy
un-matter diffusion  provided a new interpretation
of  $\sqrt\theta$ as the \emph{critical temperature} marking the transition from a smooth 
geometry
to a  trans-Planckian ``spacetime steam'' \cite{Nicolini:2010bj}. This new scenario and its
connection with the UV self-complete quantum gravity model are currently under investigation.

\ack One of us (SA) would like to acknowledge the Yukawa Institute of Kyoto University for 
partial
support during the long-term Workshop on Gravity and Cosmology (GC2010: YITP-T-10-01), 
when the foundations of this work were laid down.

\end{document}